# Observing exoplanets with the MicroObservatory: 43 new transit light curves of the hot Jupiter HAT-P-32b

Martin J. F. Fowler, Frank F. Sienkiewicz, Robert T. Zellem & Mary E. Dussault

Observations of 43 complete transits of the hot Jupiter exoplanet HAT-P-32b using the MicroObservatory 0.15m robotic telescope network, covering a period of seven years, are presented. Compared with the most recent ephemeris for the system, the precision of the mid-transit times yields a root-mean-square value from the predicted model of 3.0min. The estimated system parameters based on *EXOFAST* modelling are broadly consistent with the default parameter values listed in the NASA Exoplanet Archive. An updated orbital period of 2.15000815 ± 0.00000013d and ephemeris of 2458881.71392 ± 0.00027BJD$_{TDB}$ are consistent with recent studies of the system using larger telescopes. Using this updated ephemeris, the predicted mid-transit time for a notional observation of HAT-P-32b by the NASA JWST mission in mid-2021 is improved by 1.4min compared with the discovery ephemeris and is approximately eight times more precise. Likewise, the mid-transit time for an observation by the ESA ARIEL mission in 2020 is improved by 1.7min. Thus, observations of transiting exoplanets by MicroObservatory and other users of small telescopes can contribute to the maintenance of the ephemerides of targets for future space-based telescope missions. We also note that one of the HAT-P-32 field stars is a delta Scuti pulsating variable and that characterisation using the same observations as this study further demonstrates the utility of MicroObservatory for the observation of stellar variability, whilst simultaneously observing transiting exoplanets for ephemeris maintenance.

## Introduction

Over two decades on from the discovery of the first planet orbiting a main-sequence star other than our own Sun,[1] the number of confirmed exoplanets in the NASA Exoplanet Archive as of 2020 May 21 was 4,158, with a similar number of candidates awaiting confirmation.[2,3] The diversity of such worlds spans a much wider range of physical conditions than those in our solar system and form a continuum, from gas giants composed mostly of hydrogen, to smaller ocean planets where water may provide most of the mass, to rock-iron terrestrial worlds in some ways similar to the Earth.[4] One particular group of exoplanets with no solar-system counterpart is known as the 'hot Jupiters' class. These planets have masses comparable to, or greater than, Jupiter but orbit very close to their primary star (*i.e.*, within 0.1au). Exoplanets appear to be ubiquitous and it is estimated that for the stars that have been searched most thoroughly, *i.e.*, main-sequence dwarfs of 0.5–1.0 solar masses ($M_\odot$), the probability that a random star has a planet is of the order of unity.[5]

Under the fortuitous condition that the orbital plane of a planetary system is coincident with the observer's line of sight, exoplanets can be observed to transit their host star, leading to a periodic slight dimming of the star.[6] Known as the 'transit method', this technique has been used to great effect by ground-based observatories such as the Hungarian-made Automated Telescope Network (HATNet),[7] and the Wide Angle Search for Planets (WASP),[8] as well as the space-based missions *CoRoT*, *Kepler* and, currently, the *Transiting Exoplanet Survey Satellite* (TESS).[9] To date, *Kepler* has been responsible for the discovery of most of the known exoplanets,[10] and TESS is predicted to discover >10,000 new transiting examples.[11]

Relatively modest equipment, such as a 0.25m Schmidt–Cassegrain telescope on an equatorial mount, can be used by amateur astronomers to yield high-precision transit light curves of hot-Jupiter exoplanets.[12] The Exoplanet Transit Database (ETD), run by the Czech Astronomical Society,[13] currently includes data on >9,500 transits of over 350 exoplanets, the majority contributed by amateur observers. Amateur data from the ETD and other sources have been used to investigate transit timing variations,[14,15] as well as for the refinement of ephemerides of transiting exoplanets with high timing uncertainties.[16]

Indeed, observations of transiting exoplanets by amateurs and other users of small telescopes (≤1m), have an important role to play in the maintenance of the ephemerides of targets for future space-based telescope missions.[17,18] These missions include the NASA *James Webb Space Telescope* (JWST, 2021 launch), the ESA ARIEL mission (2028 launch), and an Astro2020 Decadal mission (~2030 launch). Without follow-up by ground-based observations, the ephemerides of many of these targets would become 'stale' by the time the missions are flown, leading to the prospect of inefficient use of valuable telescope time because of the uncertainty in the transit timings. To this end, projects such as NASA Exoplanet Watch and ARIEL ExoClock,[17,19,20,21] both of which are open to amateur observers and other users of small telescopes, have been established to monitor transiting exoplanets in order to keep their ephemerides up to date.

Observations of transiting exoplanets also form the basis of the Laboratory for the Study of Exoplanets (ExoLab), an online teaching resource that is operated by the Science Education Department at the Center for Astrophysics, Harvard & Smithsonian.[22,23] Developed with funding from the US National Science Foundation, and aimed at high-school classrooms in physics, astronomy and Earth science, ExoLab uses the 6-inch (152mm) telescopes of the MicroObservatory robotic telescope network to enable students to detect and analyse the transits of known exoplanets using rudimentary online photometry and modelling tools. Since 2009, MicroObservatory has taken >3,500 transit datasets to serve science students and interested exoplanet observers.

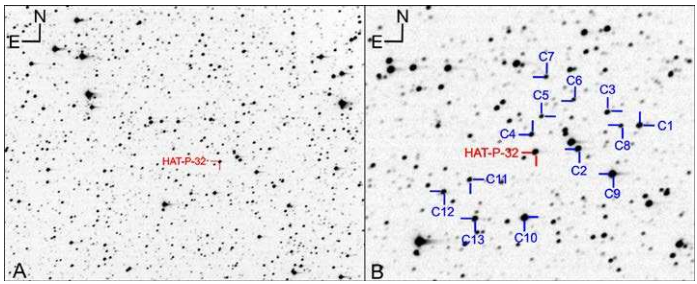

**Figure 1. (A)** Representative unfiltered full-frame MicroObservatory image of the HAT-P-32 field acquired on 2020 Jan 18. **(B)** Enlargement, showing the comparison stars used for aperture photometry (identified as C1 through C13).

Complementing ExoLab, the 'DIY Planet Search' website – a project of NASA's Universe of Learning[24,25] – is a public engagement tool that uses the same image data as ExoLab and allows anyone to investigate the transit method. Whilst a similar online photometry tool is provided on the website, the images can be downloaded as FITS files for offline reduction and analysis.[26] Importantly, these are all new observations; over the course of a year many are made of around 30 known exoplanets.

One of the targets regularly observed by DIY Planet Search is the $V$ = 11.3mag, late-F–early-G dwarf star GSC 3281-00800 (= HAT-P-32), which is orbited by the exoplanet HAT-P-32b. Discovered by the HATNet in 2004 and confirmed some seven years later, HAT-P-32b was initially found to have a mass of ~0.86$M_J$, radius of ~1.789$R_J$, transit depth of the order of 20 millimagnitudes and an orbital period, $P$, of ~2.15d.[27] The mass and radius estimates were subsequently refined to 0.68$M_J$ and 1.98$R_J$ respectively.[28] The planet's dayside temperature has been measured by Zhao *et al.* (2014) to be $T_{eq}$ = 2,042 ± 50K,[29] and its optical transmission spectrum shows little variation, suggesting a high-altitude cloud layer masking any atmospheric features.[30–33] Seeliger *et al.* (2014) have shown from transit timing analysis that the HAT-P-32 system does not exhibit transit timing variations (TTVs), arising from additional unknown bodies, of more than ~1.5min,[34] making it a good target to evaluate the accuracy and suitability of MicroObservatory observations for exoplanet research.

In this paper, we present observations of 43 transits of exoplanet HAT-P-32b by the MicroObservatory together with estimates of the parameters of the system based on the analysis of combined transit light curves, and an updated ephemeris.

## Observations

Between 2013 and 2020, 43 complete transits by HAT-P-32b were successfully observed by the MicroObservatory robotic telescopes (Table 1). During this period, observations of a further 40 transit events were attempted, of which 11 provided partial transits and the remainder were unsuccessful either because of: technical issues (10); poor weather (14); or very noisy transits that could not be analysed successfully (5). Overall, the success rate of observing complete HAT-P-32b transits was ~52%.

Except for the observations of 2014 Oct 4, which were made using the MicroObservatory telescope *Donald*, all the observations were made with the telescope *Cecilia*.[35] Located at the Smithsonian's Fred Lawrence Whipple Observatory (FLWO) in Amado, Arizona, USA, the MicroObservatory telescopes are of an original Maksutov design, with a 6-inch [152mm] spherical primary mirror and a focal length of 560mm.[22] A Kodak KAF-1402ME CCD sensor produces images covering a field of view of 0.96×0.75°, with a nominal plate scale of 5.21 arcsec/pixel (2×2 binning). For each observation run, we obtained unfiltered 60s-exposure science FITS images with a three-minute cadence, from around one hour before and one hour after the predicted start/end of the transit, giving approximately 100 images per run. The headers of the FITS images were time-stamped with the start of exposure based on a local precise time server at the FLWO and are accurate to the nearest second.

In addition to the science images, two dark-field images were collected on each run. Whilst flat-field (FF) images were not available, for the 2016 runs which were particularly noisy with dust motes, a master pseudo-FF was prepared from nine images of the exoplanetary system WASP-52, obtained on an overcast moonlit night. This pseudo-FF to some extent offset the 'wandering' of the target and comparison stars across the sensor throughout the sequence of images, although this remains a significant source of noise in the datasets. A representative science image showing the HAT-P-32 field is shown in Figure 1.

## Data reduction & analysis pipeline

We developed a partially automated MicroObservatory Exoplanet Observation Workflow pipeline (*MEOW* v. 1.0) for reduction and analysis of the DIY Planet Search images. For each set of observations, the science images, together with a dark-field image and, where available, the pseudo-FF image, were imported into *MuniWin* (v. 2.1.24 (×64)),[36] for automated differential aperture photometry using an ensemble of up to eight comparison stars as shown in Figure 1 and detailed in Table 1. The measuring aperture was typically three pixels in radius and the background sky was measured using a concentric aperture of two pixels, with a gap of one pixel between the two apertures. We then saved the photometry output from *MuniWin* (Julian Date ($JD_{UTC}$) of mid-observation, $V$–$C$ magnitude, error), together with the airmass, as text files and uploaded the photometry output to the ETD to gain an initial indication of the quality of the transit (Level 1 analysis).

For more detailed analysis, we imported the text file outputs from *MuniWin* into a custom Microsoft *Excel* spreadsheet for conversion into a format suitable for use with the *EXOFAST* transit-fitting model developed by Eastman *et al.* (2013).[37] For this higher-fidelity analysis (Level 2), the $V$–$C$ magnitudes were converted to relative fluxes and the time converted from $JD_{UTC}$ to Barycentric Julian Date in Barycentric Dynamical Time ($BJD_{TDB}$) using the online utility developed by Eastman *et al.* (2010).[38] For each data point, a linear de-trend parameter was calculated based on the slope of the pre- and post-transit data points as identified from the Level 1 analysis, and the fluxes normalised to an out-of-transit value of ~1. We then prepared an 'output' text file comprising: $BJD_{TDB}$, normalised flux, flux error, linear de-trend parameter and airmass. This file was used as the photometry input to the *EXOFAST* transit-fitting model.

A full demonstration of the use of *MuniWin* software with MicroObservatory exoplanet image data is available on YouTube.[39]

## Transit-fitting modelling

*EXOFAST* has become an important tool for astronomers who want to use transit light curves or radial velocity data, or both, together with various inputs, to create models of planetary systems. Originally requiring the use of the proprietary software language IDL, the NASA Exoplanet Archive has recently integrated the same IDL-based calculations as the original into its suite of web resources.[3] The software implements the light curve models of Mandel & Agol (2002),[40] as well as a differential-evolution Markov Chain Monte Carlo (MCMC) method,[41,42] to estimate and characterise parameters and uncertainty distributions.[43] The Exoplanet Archive website provides enough back-end computing resources to enable the MCMC analysis of observed transit light curves and thus relieves the user of the need to run this high-fidelity model locally.

We uploaded individual output photometry files from the *MEOW* pipeline to *EXOFAST*, and the default parameter set for HAT-P-32b required for the model was 'pulled' from the Exoplanet Archive. The various prior values of the required parameters, hereafter termed 'priors', together with their widths (uncertainties) are given in Table 2. These priors represent the default parameter values for the system in late 2016 and whilst they have been updated since then, these have been used in the analysis to maintain continuity across the observation datasets. Limb-darkening coefficients were automatically calculated by *EXOFAST* for each run. Except for the run using combined transits covering multiple epochs, a mid-transit ($T_C$) prior was not specified since for transit-only fits the mean of the input times is used by the model as the prior when no midpoint is specified.

With transit-only fits, the light-curve data alone provide very little constraint on the eccentricity and longitude of periastron, both of which are required for the model. These parameters appreciably affect the derived physical parameters. Accordingly, we forced a circular orbit for the transit fits, as recommended by the *EXOFAST* documentation. Such an orbit is consistent with the near-circular solution of Zhao *et al.* (2014),[29] although others have derived eccentricities in the range 0.10–0.22.[27,28]

Since *EXOFAST* does not include a generic 'clear' filter for the modelling of transit light curves, we chose the *CoRoT* band as it approximates to the 'clear' filter used for the MicroObservatory images.[44]

**Table 1. Journal of observations for 43 light curves of HAT-P-32b transits**

| Date (UT) | Epoch | Telescope | Observations | Airmass | Comparison stars |
|---|---|---|---|---|---|
| 2013 Feb 6 | 215 | *Cecilia* | 71 | 1.1→2.5 | C1→C8 |
| 2013 Sep 24 | 322 | *Cecilia* | 91 | 3.3→1.1 | C1→C8 |
| 2013 Sep 26 | 323 | *Cecilia* | 97 | 1.3→1.0→1.1 | C1→C8 |
| 2013 Oct 9 | 329 | *Cecilia* | 99 | 1.7→1.0 | C1→C8 |
| 2013 Oct 11 | 330 | *Cecilia* | 87 | 1.0→1.3 | C1→C8 |
| 2013 Oct 24 | 336 | *Cecilia* | 94 | 1.2→1.0→1.2 | C1→C8 |
| 2013 Nov 6 | 342 | *Cecilia* | 97 | 1.5→1.0 | C1→C8 |
| 2013 Nov 8 | 343 | *Cecilia* | 92 | 1.0→1.7 | C1→C8 |
| 2013 Dec 4 | 355 | *Cecilia* | 92 | 1.3→1.0→1.1 | C1→C8 |
| 2014 Jan 3 | 369 | *Cecilia* | 91 | 1.0→2.7 | C1→C8 |
| 2014 Jan 16 | 375 | *Cecilia* | 91 | 1.0→1.7 | C1→C8 |
| 2014 Oct 4 | 496 | *Donald* | 103 | 1.4→1.0 | C1→C8 |
| 2014 Oct 18 | 503 | *Cecilia* | 99 | 1.1→1.0→1.3 | C1→C8 |
| 2014 Oct 31 | 509 | *Cecilia* | 77 | 1.3→1.0→1.1 | C1→C8 |
| 2014 Nov 15 | 516 | *Cecilia* | 88 | 1.0→1.5 | C1→C8 |
| 2014 Nov 28 | 522 | *Cecilia* | 95 | 1.2→1.0→1.1 | C1→C8 |
| 2015 Oct 23 | 675 | *Cecilia* | 86 | 2.4→1.0 | C1→C8 |
| 2015 Nov 7 | 682 | *Cecilia* | 102 | 1.4→1.0 | C1→C8 |
| 2015 Nov 9 | 683 | *Cecilia* | 89 | 1.0→1.8 | C1→C8 |
| 2016 Sep 5 | 823 | *Cecilia* | 91 | 1.6→1.0 | C1→C8 |
| 2016 Oct 3 | 836 | *Cecilia* | 89 | 1.6→1.0→1.1 | C1→C8 |
| 2016 Oct 18 | 843 | *Cecilia* | 121 | 1.2→1.0→1.5 | C1→C8 |
| 2016 Oct 31 | 849 | *Cecilia* | 117 | 1.5→1.0→1.2 | C1→C8 |
| 2017 Sep 25 | 1002 | *Cecilia* | 98 | 3.6→1.1 | C4, C9→C13 |
| 2017 Sep 27 | 1003 | *Cecilia* | 96 | 1.2→1.0→1.1 | C1→C8 |
| 2017 Oct 10 | 1009 | *Cecilia* | 100 | 1.6→1.0 | C1→C8 |
| 2017 Oct 23 | 1015 | *Cecilia* | 95 | 2.3→1.0 | C1→C8 |
| 2017 Nov 7 | 1022 | *Cecilia* | 95 | 1.5→1.0→1.1 | C2→C4 |
| 2017 Nov 22 | 1029 | *Cecilia* | 104 | 1.1→1.0→1.4 | C1→C8 |
| 2017 Dec 20 | 1042 | *Cecilia* | 122 | 1.1→1.0→1.7 | C1→C8 |
| 2018 Nov 1 | 1189 | *Cecilia* | 101 | 1.2→1.0→1.1 | C1→C8 |
| 2018 Nov 14 | 1195 | *Cecilia* | 95 | 1.7→1.0 | C1→C8 |
| 2018 Nov 29 | 1202 | *Cecilia* | 90 | 1.2→1.0→1.1 | C1→C8 |
| 2018 Dec 14 | 1209 | *Cecilia* | 93 | 1.0→1.6 | C1→C8 |
| 2019 Jan 11 | 1222 | *Cecilia* | 92 | 1.0→2.2 | C1→C8 |
| 2019 Sep 13 | 1336 | *Cecilia* | 91 | 1.8→1.0 | C1→C4 |
| 2019 Sep 28 | 1343 | *Cecilia* | 104 | 1.2→1.0→1.1 | C1 |
| 2019 Oct 24 | 1355 | *Cecilia* | 102 | 2.5→1.0 | C1→C4 |
| 2019 Dec 6 | 1375 | *Cecilia* | 103 | 1.3→1.0→1.1 | C1→C4 |
| 2019 Dec 21 | 1382 | *Cecilia* | 102 | 1.0→1.5 | C1 |
| 2020 Jan 5 | 1389 | *Cecilia* | 100 | 1.0→3.0 | C1→C4 |
| 2020 Jan 18 | 1395 | *Cecilia* | 103 | 1.0→1.8 | C1→C4 |
| 2020 Feb 2 | 1402 | *Cecilia* | 85 | 1.1→4.2 | C1→C4 |

**Table 2. HAT-P-32b prior value & width inputs for *EXOFAST* modelling**

| Prior parameter | Prior value | Prior width |
|---|---|---|
| Orbital inclination, $i$ (degrees) | 88.9 | – |
| Stellar effective temp., $T_{eff}$ (Kelvin) | 6269 | 64.0 |
| Metallicity, [Fe/H] (dex) | –0.04 | 0.08 |
| Orbital period, $P$ (d) | 2.15000805 | 0.00000095 |
| Eccentricity, $e$ | 0.0 | – |
| Longitude of periastron, $\omega$ * (degrees) | 90.0 | – |

MCMC fits were successfully run on 28 of the 43 transit light curves. The failures of the remaining 15 light curves were mainly because of the presence of NaN ('not a number') errors, which caused the EXOFAST_PLOTDIST routine to halt and no output to be produced. Unfortunately, this could not be addressed because of the reduced input parameters of the flavour of EXOFAST run on the Exoplanet Archive website compared with the full offline version. For these transits, chi-squared (chi2) fits were successfully made using the model, although it should be noted that the uncertainties of the derived parameters from these fits are analytical approximations rather than the more accurate statistical estimations determined by the MCMC fits.[45]

*Fowler et al.: 43 new transit light curves of the hot Jupiter HAT-P-32b*

## Analysis of transit light curves

A representative model fit (with residuals) to the observed light curve of the HAT-P-32b transit observed on 2019 Dec 6 (epoch 1375) is shown in Figure 2. All 43 light curves, together with their respective modelled transit curves, are shown in Figure 3. The light curves show well-defined transits, albeit with varying degrees of scatter around the transit curves. Key derived parameters, including median times of mid-transit ($T_C$), total duration (1st to 4th contact, $T_{14}$) and transit depth, are tabulated for each transit in Table 3.

### Mid-transit times

Wang *et al.* (2019) have recently refined the linear ephemeris of the HAT-P-32b system to be:[28]

$$T_C(N) = 2455867.402743\ (49) + N \cdot 2.15000820\ (13)$$

where $N$ represents the number of orbital cycles (epochs) since the reference epoch (given in $BJD_{TDB}$) and the bracketed

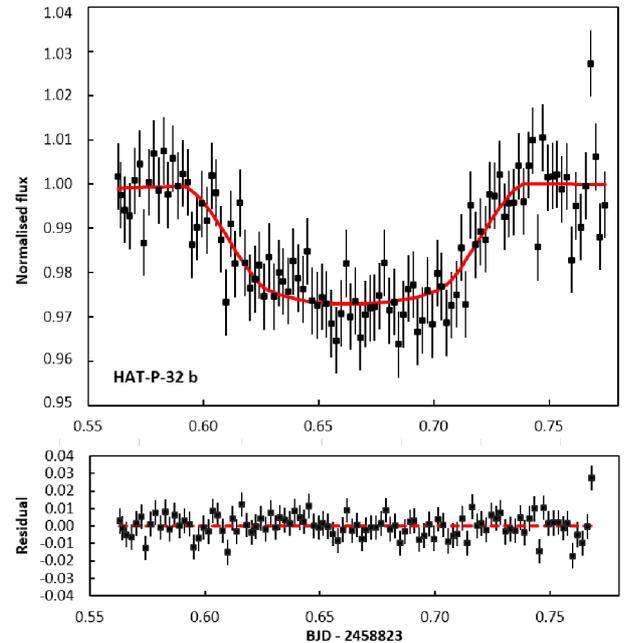

**Figure 2.** Representative transit light curve of HAT-P-32b for observations of 2019 Dec 6 (epoch 1375), together with the modelled transit fit using *EXOFAST*.

### Table 3. Derived parameters & 68% confidence intervals for 43 transits of HAT-P-32b

| Date (UT) | Epoch | Fit type | RMS residuals | Time of mid-transit, $T_C$ ($BJD_{TDB}$) | O–C (min) | Total duration, $T_{14}$ (min) | Depth (%) |
|---|---|---|---|---|---|---|---|
| 2013 Feb 6 | 215 | Chi2 | 0.007 | 2456329.6556 ±0.0019 | 1.5 | 195.6 ±10.8 | 2.2 ±0.1 |
| 2013 Sep 24 | 322 | Chi2 | 0.010 | 2456559.7043 ±0.0017 | –1.5 | 180.6 ±9.6 | 3.0 ±0.2 |
| 2013 Sep 26 | 323 | MCMC | 0.007 | 2456561.8566 ±0.0026 | 1.8 | 193.1 ±17.9 | 2.5 ±0.6 |
| 2013 Oct 9 | 329 | MCMC | 0.007 | 2456574.7597 ±0.0024 | 6.1 | 213.9 ±24.8 | 2.9 ±0.8 |
| 2013 Oct 11 | 330 | MCMC | 0.007 | 2456576.9060 ±0.0019 | 0.8 | 200.3 ±9.9 | 3.0 ±0.8 |
| 2013 Oct 24 | 336 | Chi2 | 0.008 | 2456589.8052 ±0.0017 | –0.4 | 186.7 ±9.8 | 2.1 ±0.1 |
| 2013 Nov 6 | 342 | Chi2 | 0.008 | 2456602.7079 ±0.0017 | 3.3 | 180.4 ±9.7 | 2.8 ±0.1 |
| 2013 Nov 8 | 343 | MCMC | 0.007 | 2456604.8559 ±0.0014 | 0.5 | 188.6 ±6.0 | 2.6 ±0.3 |
| 2013 Dec 4 | 355 | Chi2 | 0.008 | 2456630.6559 ±0.0021 | 0.4 | 216.8 ±12.0 | 2.5 ±0.1 |
| 2014 Jan 3 | 369 | Chi2 | 0.007 | 2456660.7537 ±0.0013 | –2.9 | 190.2 ±16.1 | 2.4 ±0.1 |
| 2014 Jan 16 | 375 | Chi2 | 0.010 | 2456673.6560 ±0.0022 | 0.2 | 178.4 ±12.4 | 2.2 ±0.2 |
| 2014 Oct 4 | 496 | MCMC | 0.004 | 2456933.8061 ±0.0009 | –1.1 | 191.4 ±6.1 | 2.6 ±0.2 |
| 2014 Oct 18 | 503 | MCMC | 0.005 | 2456948.8583 ±0.0022 | 2.1 | 203.1 ±31.9 | 2.8 ±1.0 |
| 2014 Oct 31 | 509 | MCMC | 0.005 | 2456961.7574 ±0.0024 | 0.7 | 193.3 ±34.2 | 2.3 ±71.4 |
| 2014 Nov 15 | 516 | MCMC | 0.006 | 2456976.8059 ±0.0023 | –1.5 | 194.7 ±33.1 | 2.5 ±16.4 |
| 2014 Nov 28 | 522 | MCMC | 0.005 | 2456989.7067 ±0.0012 | –0.4 | 210.2 ±11.3 | 3.1 ±0.4 |
| 2015 Oct 23 | 675 | MCMC | 0.006 | 2457318.6592 ±0.0013 | 1.3 | 199.6 ±8.0 | 2.9 ±0.4 |
| 2015 Nov 7 | 682 | MCMC | 0.006 | 2457333.7105 ±0.0022 | 3.2 | 162.8 ±8.5 | 2.2 ±0.9 |
| 2015 Nov 9 | 683 | MCMC | 0.006 | 2457335.8567 ±0.0025 | –2.3 | 193.8 ±25.3 | 2.4 ±0.6 |
| 2016 Sep 5 | 823 | MCMC | 0.007 | 2457636.8569 ±0.0018 | -3.7 | 215.1 ±13.7 | 2.7 ±0.6 |
| 2016 Oct 3 | 836 | MCMC | 0.007 | 2457664.8105 ±0.0021 | 1.4 | 193.4 ±11.2 | 2.1 ±0.3 |
| 2016 Oct 18 | 843 | MCMC | 0.007 | 2457679.8601 ±0.0016 | 0.6 | 192.2 ±11.4 | 2.8 ±0.3 |
| 2016 Oct 31 | 849 | MCMC | 0.007 | 2457692.7620 ±0.0017 | 3.3 | 198.9 ±9.2 | 2.9 ±0.3 |
| 2017 Sep 25 | 1002 | Chi2 | 0.009 | 2458021.7084 ±0.0022 | –3.7 | 186.6 ±12.4 | 1.9 ±0.1 |
| 2017 Sep 27 | 1003 | MCMC | 0.005 | 2458023.8619 ±0.0012 | 1.4 | 188.0 ±9.2 | 2.8 ±0.3 |
| 2017 Oct 10 | 1009 | MCMC | 0.005 | 2458036.7623 ±0.0012 | 1.9 | 195.1 ±9.8 | 2.4 ±0.3 |
| 2017 Oct 23 | 1015 | Chi2 | 0.007 | 2458049.6618 ±0.0011 | 1.1 | 188.8 ±6.6 | 2.5 ±0.1 |
| 2017 Nov 7 | 1022 | MCMC | 0.008 | 2458064.7114 ±0.0025 | 0.4 | 198.3 ±18.0 | 2.4 ±0.4 |
| 2017 Nov 22 | 1029 | MCMC | 0.006 | 2458079.7598 ±0.0019 | –2.0 | 192.2 ±13.0 | 2.5 ±0.3 |
| 2017 Dec 20 | 1042 | MCMC | 0.006 | 2458107.7118 ±0.0014 | 0.7 | 176.2 ±6.1 | 2.2 ±0.2 |
| 2018 Nov 1 | 1189 | MCMC | 0.007 | 2458423.7592 ±0.0015 | –4.7 | 185.9 ±5.7 | 2.0 ±0.2 |
| 2018 Nov 14 | 1195 | Chi2 | 0.013 | 2458436.6638 ±0.0022 | 1.7 | 196.8 ±12.5 | 2.9 ±0.2 |
| 2018 Nov 29 | 1202 | MCMC | 0.007 | 2458451.7056 ±0.0016 | –10.1 | 174.3 ±7.3 | 2.0 ±0.2 |
| 2018 Dec 14 | 1209 | Chi2 | 0.010 | 2458466.7637 ±0.0025 | 1.5 | 179.5 ±14.3 | 1.8 ±0.2 |
| 2019 Jan 11 | 1222 | MCMC | 0.009 | 2458494.7074 ±0.0018 | –7.8 | 187.5 ±11.7 | 2.4 ±0.3 |
| 2019 Sep 13 | 1336 | Chi2 | 0.008 | 2458739.8133 ±0.003 | –0.5 | 207.8 ±17.1 | 1.4 ±0.1 |
| 2019 Sep 28 | 1343 | MCMC | 0.007 | 2458754.8637 ±0.0026 | –0.1 | 195.3 ±5.6 | 3.9 ±3.9 |
| 2019 Oct 24 | 1355 | MCMC | 0.007 | 2458780.6648 ±0.002 | 1.4 | 186.9 ±11.1 | 1.7 ±0.2 |
| 2019 Dec 6 | 1375 | MCMC | 0.007 | 2458823.6659 ±0.0025 | 2.8 | 216.3 ±30.9 | 2.6 ±16.7 |
| 2019 Dec 21 | 1382 | MCMC | 0.011 | 2458838.7139 ±0.0017 | –0.3 | 172.9 ±16.6 | 2.2 ±0.5 |
| 2020 Jan 5 | 1389 | Chi2 | 0.011 | 2458853.7616 ±0.0026 | –3.7 | 181.1 ±15.2 | 1.9 ±0.2 |
| 2020 Jan 18 | 1395 | MCMC | 0.008 | 2458866.6615 ±0.0021 | –3.9 | 187.9 ±12.3 | 2.3 ±0.3 |
| 2020 Feb 2 | 1402 | Chi2 | 0.011 | 2458881.7117 ±0.0027 | –3.7 | 184.2 ±15.3 | 2.0 ±0.2 |

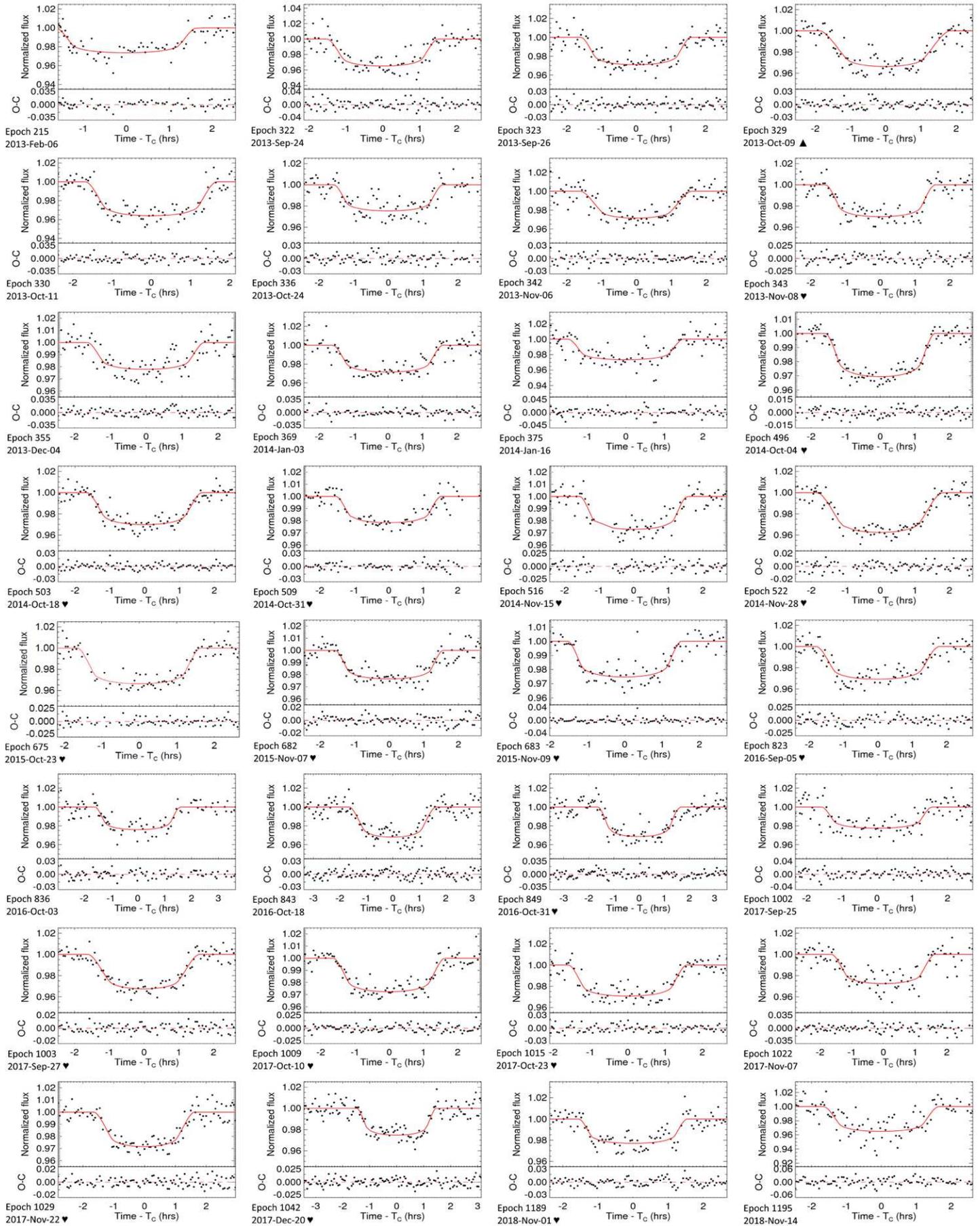

**Figure 3.** Transit light curves and *EXOFAST* model fits for 43 transits of HAT-P-32b observed by MicroObservatory (figure continued next page). The three transits exhibiting $T_C$ *O–C* values in excess of ±6min are annotated with a triangle (▲) after the transit date. Likewise, the top 20 transits used to determine the HAT-P-32b system parameters are annotated with a heart (♥).

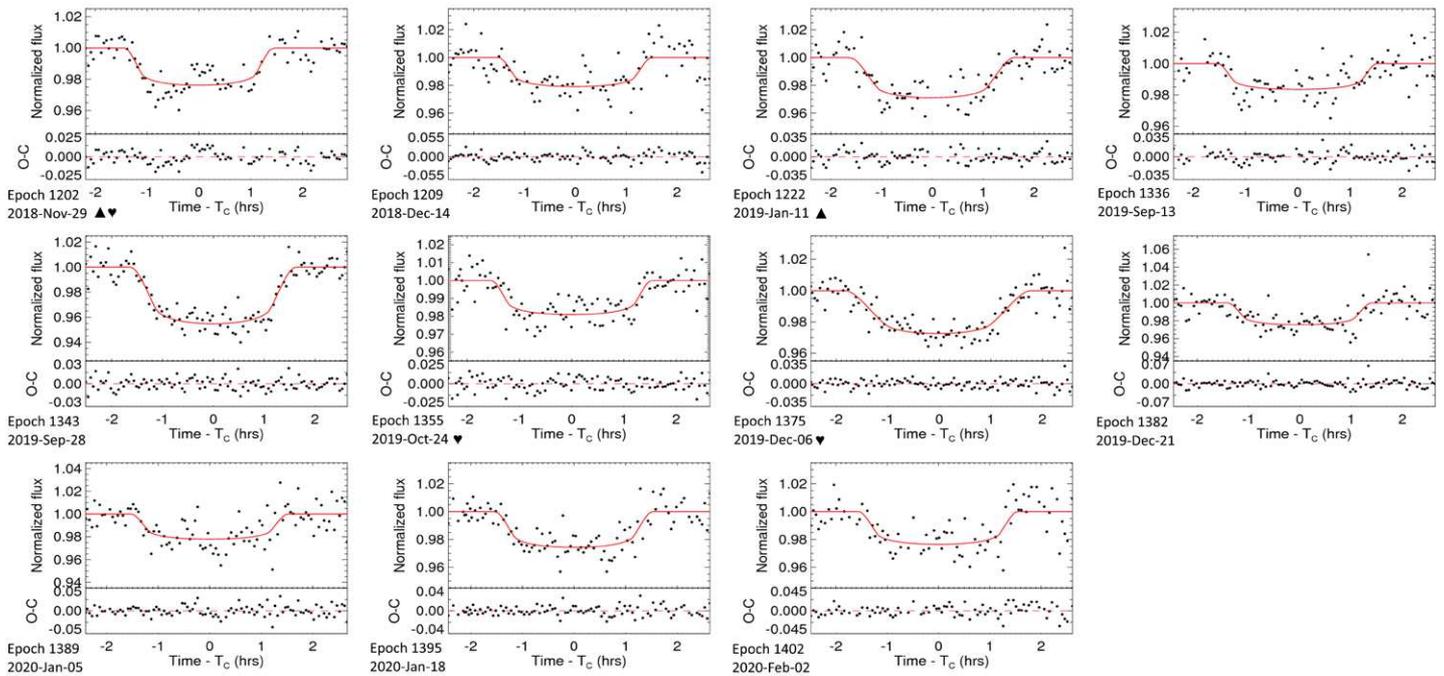

quantities represent the uncertainty in the final two digits of the preceding number.

A plot of the observed-minus-calculated (*O–C*) values for $T_C$ based on this ephemeris is shown in Figure 4. The mean of *O–C* values is –0.3 ± 3.0min and the scatter of all the mid-transit times has a root mean square (RMS) of 3.0min. The reduced chi-squared statistic ($\chi^2_\nu$) is 1.02 with 42 degrees of freedom, indicating that the observed mid-transit times agree with the calculated times at the one-sigma level.

Three of the transits have $T_C$ *O–C* values that are in excess of ±6min. The transit of 2013 Oct 9 (epoch 329) has a $T_C$ that is 6.1min late, whereas the transits of 2018 Nov 29 (epoch 1202) and 2019 Jan 11 (epoch 1222) are respectively 10.1 and 7.8min early. Whilst these *could* reflect TTVs due to another unknown body in the system, the transit light curves for these sets of observations are relatively noisy (as highlighted in Figure 3) and in the latter two cases the transits are somewhat poorly defined, suggesting that they are likely to be artefacts rather than actual TTVs.

## Transit duration

A histogram of the total transit durations (1st to 4th contact, $T_{14}$) of the 43 observed transits is shown in Figure 5. The weighted mean transit duration is 189.1 ± 12min and this compares well with the estimate of Wang *et al.* of 187.2 ± 0.8, although the error of the MicroObservatory observations is over an order of magnitude greater and reflects the greater noise of the transit observations.

## Transit depth

Likewise, a histogram of the transit depth of the 43 observed transits is shown in Figure 6. The weighted mean transit depth is 2.28 ± 0.45% and is comparable to the depth of 2.216 ± 0.017% reported by Wang *et al.*, who used a red filter for their observations rather than unfiltered observations as used in this study.

## Estimating the HAT-P-32b system parameters

We estimated the HAT-P-32b system parameters from the current observations by running the combined photometry from the top 20 transits, as determined by their individual RMS residuals, in *EXOFAST*. Priors and widths were as before except that in this case a $T_C$ prior of 2457692.7597 ± 0.0020BJD$_{TDB}$ was used, as this equates to the predicted $T_C$ for the median transit of the set (2016 Oct 31, epoch 849), based on the Wang *et al.* ephemeris given above.

The resulting combined transit light curve of the 1,952 data points is shown in Figure 7, with the median values and 68% confidence intervals for the system parameters given in Table 4 (p.367). As can be seen, the estimated system parameters are broadly consistent with those of Wang *et al.*, which were the default parameter values listed in the NASA Exoplanet Archive for the HAT-P-32b system at the time of writing.

## Calculating an updated ephemeris

Using the derived mid-transit times and uncertainties for our 43 observed transits of HAT-P-32b, we can calculate an updated ephemeris to predict future transit events more accurately. To illustrate the power of follow-up observations to refine the original ephemeris, we use the original discovery ephemeris of Hartman *et al.* (2011) as priors for the system,[27] and solve for HAT-P-32b's orbital period and ephemeris with a MCMC fit to our MicroObservatory results,[41] following the procedure explained, for example, in detail by Zellem *et al.* (2020).[18] From this we find an

orbital period of 2.15000711 ± 0.00000057d and ephemeris of 2458881.71327 ± 0.00044BJD$_{TDB}$. Likewise, to illustrate the value of follow-up observations even with existing high-precision data, we use the default parameter values in the NASA Exoplanet Archive (*i.e.*, Wang *et al.*[28]) as priors to the MCMC fit, to find an orbital period of 2.15000815 ± 0.00000013d and ephemeris of 2458881.71392 ± 0.00027BJD$_{TDB}$.

Taking these updated ephemerides and propagating the errors forward as described in Zellem *et al.*,[18] we can estimate the $T_C$ for a NASA JWST mission in mid-2021. When we use the discovery ephemeris as priors to the MCMC fit, we calculate the $T_C$ of a notional transit of HAT-P-32b on 2021 May 30 to be like that using the discovery ephemeris alone (Figure 8A). However, it is ~4 times more precise and thus shows the benefit of the follow-up observations to refine the precision of the original ephemeris. When we use the NASA Exoplanet Archive ephemeris as the priors to the MCMC fit, the calculated $T_C$ is 1.4min later and ~8 times more precise than that calculated using the discovery ephemeris, which at that time would be over 13 years old. Indeed, this estimate is only slightly less accurate and 0.5min earlier than that calculated using the ephemeris given in the NASA Exoplanet Archive and shows the value of follow-up observations, even with existing high-precision data.

In the case of an ESA ARIEL mission in 2028, we find that the shorter orbital period determined by the MCMC fit using the discovery ephemeris as priors gives an estimated $T_C$ that is 1.3min earlier than that using the discovery ephemeris alone. However, when we use the MCMC fit with the NASA Exoplanet Archive ephemeris as the prior, we find a $T_C$ that is consistent with that obtained using the NASA Exoplanet Archive and 1.7min later than that predicted using the discovery ephemeris (Figure 8B).

On this basis, we propose an updated ephemeris of the HAT-P-32b system of:

$$T_C(N) = 2458881.71392\ (27) + N \cdot 2.15000815\ (13)$$

where, again, the bracketed quantities represent the uncertainty in the final two digits of the preceding number. Whilst there is a small difference in the period compared with that determined using the *EXOFAST* analysis of the top 20 transit curves (see Table 4), it is nevertheless contained within the larger uncertainty of the *EXOFAST* estimate.

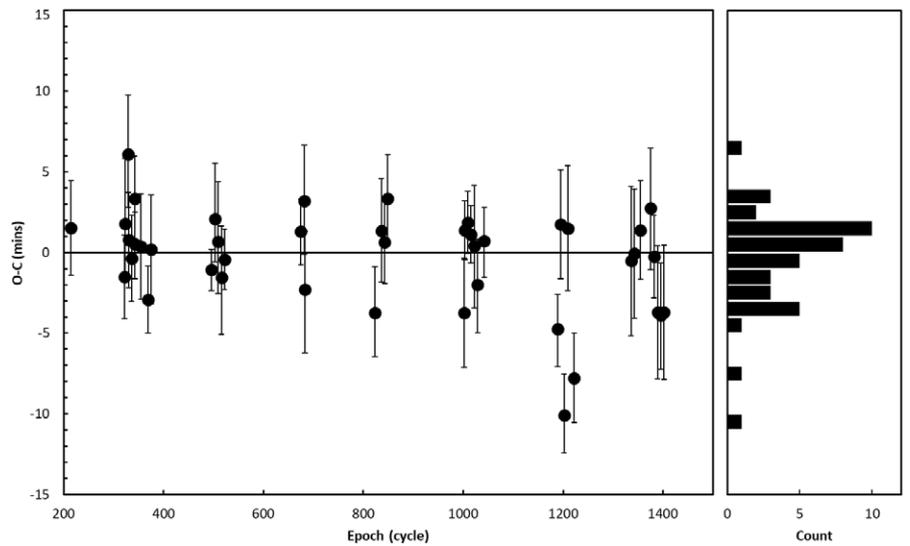

**Figure 4.** Observed-minus-calculated (*O*–*C*) mid-transit ($T_C$) residuals for the 43 observed transits of HAT-P-32b, together with a histogram showing the distribution of the individual values. The $T_C$ residuals were calculated using the ephemeris of Wang *et al* (2019).

## Discussion & conclusions

In this work we have presented 43 complete transit light curves of the hot Jupiter HAT-P-32b, acquired by the MicroObservatory robotic telescope network. Compared with the most recent estimate of the ephemeris of the system by Wang *et al.*,[28] the mean of the $T_C$ *O*–*C* values is –0.3 ± 3.0min and the scatter of all the mid-transit times has an RMS of 3.0min, which is similar to the nominal cadence of the observations. Although three of the transits show $T_C$ *O*–*C* values that are greater than ±6min, the transit light curves are quite noisy, with two of the transits being somewhat poorly defined. Since Seeliger *et al.* have shown that transit timing analysis of the HAT-P-32 system excludes TTVs of more than ~1.5min,[34] these outlier points on the *O*–*C* plot are considered to be artefacts rather than actual TTVs.

The mean total transit duration ($T_{14}$) and transit depth are comparable with the Wang *et al.* estimates. Similarly, the estimated system parameters based on *EXOFAST* modelling of the combined top 20 transit light curves are broadly consistent with those of Wang *et al*. These represent the default parameter values listed in the NASA Exoplanet Archive and indicate that observations using MicroObservatory can be used to characterise exoplanet systems to a reasonable degree of accuracy.

As noted above, observations of transiting exoplanets by amateurs,

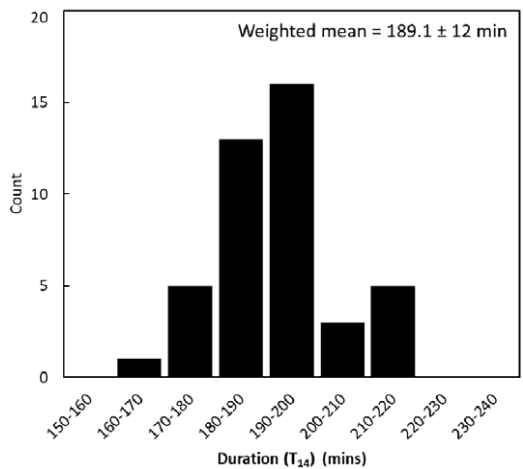

**Figure 5.** Distribution of transit durations (1st to 4th contact, $T_{14}$) for the 43 observed transits of HAT-P-32b.

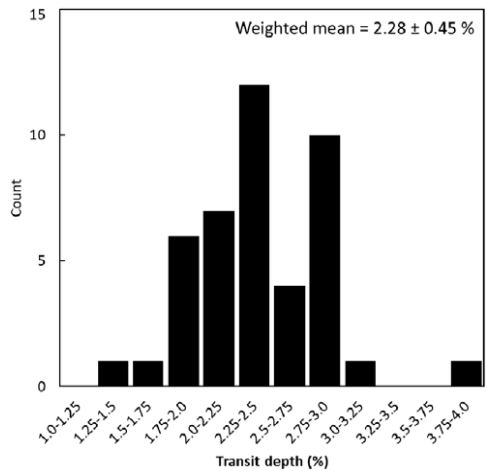

**Figure 6.** Distribution of transit depths for the 43 observed transits of HAT-P-32b.

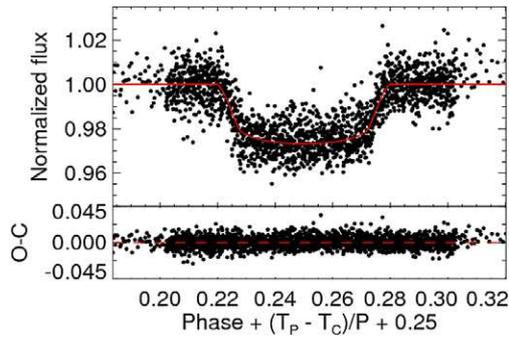

**Figure 7.** Combined transit light curve and *EXOFAST*-modelled transit based on the observations of 20 transits of HAT-P-32b. The x-axis is essentially the phase offset, so that the mid-transit occurs at 0.25.

and other users of small (≤1m) telescopes, can contribute to the maintenance of the ephemerides of targets for future space-based telescope missions that will investigate exoplanets.[17,18] Our updated ephemeris improves the predicted $T_C$ for an observation of HAT-P-32b by a JWST mission in mid-2021 by 1.4min compared with the discovery ephemeris and is ~8 times more precise. The prediction is slightly less accurate and 0.5min earlier than that calculated using the ephemeris in the NASA Exoplanet Archive. Likewise, when we use our updated ephemeris for an observation of HAT-P-32b by the ARIEL mission in 2028, we find a $T_C$ that is 1.7min later than that predicted using the discovery ephemeris and consistent with that obtained using the NASA Exoplanet Archive alone. Thus, our HAT-P-32b observations can be seen to significantly improve the original ephemeris of the system to a level consistent with recent studies using larger telescopes. Furthermore, we anticipate that observations using MicroObservatory and other small telescopes could be used to substantially refine the ephemerides of targets with larger uncertainties.

The current 30 exoplanet targets that are routinely observed by MicroObservatory throughout the year have V-band magnitudes in the range 10.3 to 14.1 and depths of transit between 1.5 and 2.9%. For 2019, we find the annual utilisation of the MicroObservatory telescope *Cecilia* in observing these targets was ~77%. Since the telescope is dedicated to observing exoplanet transits, the spare observing capacity, equivalent to ~84 nights of observations per year, could be directed towards the observation of new targets that are within the capabilities of the telescope, including some of those identified as being ideal for TTV measurements by small ground-based observatories.[18]

With a field of view of approximately 0.72deg,[2] the HAT-P-32 images show over 400 discrete stars, including the 13.9 V-magnitude δ Scuti variable star UCAC4 686-012519 (= ASASSN-V J020549.64+470040.9), which has been further characterised using the same observations as our study.[46] This finding demonstrates the potential for MicroObservatory to observe stellar variability, whilst simultaneously observing transiting exoplanets for ephemeris maintenance.[18]

## Acknowledgements

We thank Jason Eastman for his advice on running *EXOFAST* and for commenting on an early version of this paper. The paper makes use of *EXOFAST* (Eastman *et al.*, 2013) as provided by the NASA Exoplanet Archive, which is operated by the California Institute of Technology, under contract with the National Aeronautics and Space Administration under the Exoplanet Exploration Program.

MicroObservatory is maintained and operated as an educational service by the Center for Astrophysics, Harvard & Smithsonian and is a project of NASA's Universe of Learning, supported by NASA award no. NNX16AC65A. Additional MicroObservatory sponsors include the National Science Foundation, NASA, the Arthur Vining Davis Foundations, Harvard University and the Smithsonian Institution. This research has made use of NASA's Astrophysical Data System (ADS) and the NASA Exoplanet Archive. The paper makes use of data tools and/or products from Exoplanet Watch, a citizen science project managed by NASA's Jet Propulsion Laboratory on behalf of NASA's Universe of Learning. This work is supported by NASA under award no. NNX16AC65A to the Space Telescope Science Institute.

Part of the research was carried out at the Jet Propulsion Laboratory, California Institute of Technology, under contract with the National Aeronautics and Space Administration. We thank Richard Miles and Roger Dymock for helpful comments that have improved the paper.

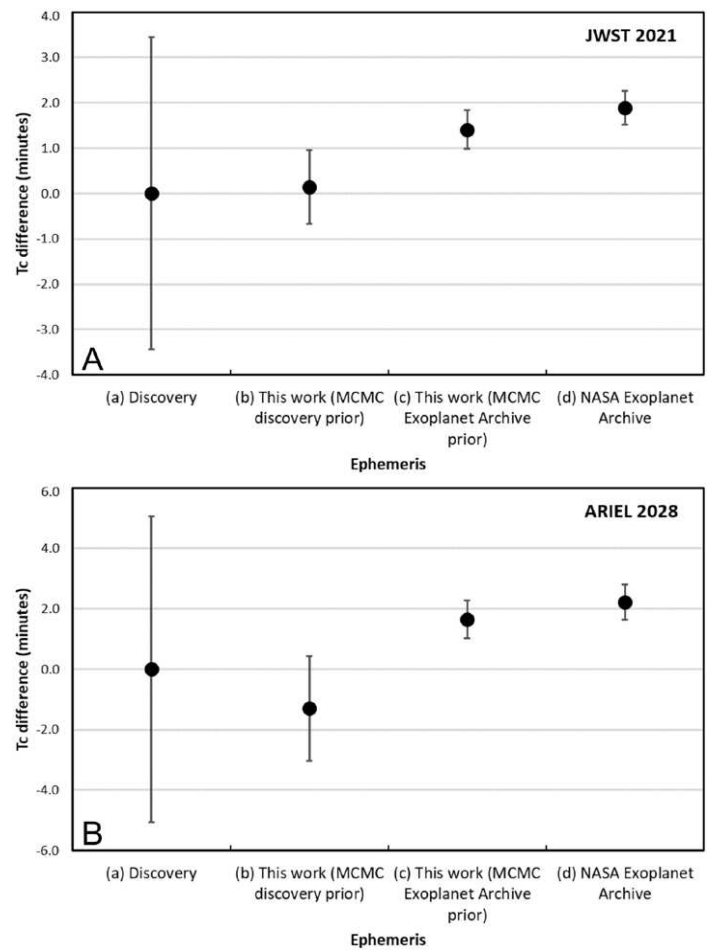

**Figure 8.** Comparison of estimated mid-transit times ($T_C$) for the observation of a notional HAT-P-32b transit by **(A)** the NASA JWST mission in mid-2021 and **(B)** the ESA ARIEL mission in 2028. Different ephemerides are used to estimate $T_C$: (a) the original discovery ephemeris; (b) an MCMC fit to the observations reported in this work using the discovery ephemeris as the prior; (c) a similar MCMC fit but using the NASA Exoplanet Archive ephemeris as the prior; (d) the NASA Exoplanet Archive ephemeris. The differences in $T_C$ are shown relative to the $T_C$ value estimated using the discovery ephemeris (*i.e.*, 2459365.464770 ± 0.00239BJD$_{TDB}$ for the JWST and 2461771.323722 ± 0.00351BJD$_{TDB}$ for ARIEL).

**Addresses**

**MJFF**: Les Rocquettes, Orchard Road, South Wonston, Winchester, SO21 3EX, UK *[danebury216@hotmail.co.uk]*. **FFS** and **MED**: Science Education Department, The Center for Astrophysics | Harvard & Smithsonian, 60 Garden Street, Cambridge, Massachusetts, 02138, USA *[ffsienkiewicz@cfa.harvard.edu; mdussault@cfa.harvard.edu]*. **RTZ**: Jet Propulsion Laboratory, California Institute of Technology, 4800 Oak Grove Drive, Pasadena, California, 91109, USA *[robert.t.zellem@jpl.nasa.gov]*.

## References & notes

1. Mayor M. & Queloz D., 'A Jupiter-mass companion to a solar-type star', *Nature*, **378**, 355–359 (1995)
2. Akeson R. L. *et al.*, 'The NASA Exoplanet Archive: data and tools for exoplanet research', *Publ. Astron. Soc. Pacific*, **125**, 989–999 (2013)
3. NASA Exoplanet Archive (2020). Available at: **https://exoplanetarchive.ipac.caltech.edu/index.html** (accessed 2020 April).
4. Spiegel D. S., Fortney J. J. & Sotin C., 'Structure of exoplanets', *Proc. Natl. Acad. Sci.*, **111**, 12622–12627 (2014)
5. Winn J. N. & Fabrycky D. C., 'The occurrence and architecture of exoplanetary systems', *Annu. Rev. Astron. Astrophys.*, **53**, 409–447 (2014)
6. Haswell C. A., *Transiting exoplanets*, Cambridge University Press, 2010
7. Bakos G. Á., 'The HATNet and HATSouth exoplanet surveys', *arXiv Prepr.*, **1801.00849**, 1–11 (2018)
8. Pollacco D. L. *et al.*, 'The WASP Project and the SuperWASP cameras', *Publ. Astron. Soc. Pacific*, **118**, 1407–1418 (2006)
9. Perryman M., *The exoplanet handbook*, 2nd edn, Cambridge University Press, 2018
10. Lissauer J. J., Dawson R. I. & Tremaine S., 'Advances in exoplanet science from *Kepler*', *Nature*, **513**, 336–344 (2014)
11. Barclay T., Pepper J. & Quintana E. V., 'A revised exoplanet yield from the *Transiting Exoplanet Survey Satellite* (TESS)', *Astrophys. J. Suppl. Ser.*, **239**(2), 15pp (2018)
12. Sergison D., 'High precision photometry: detection of exoplanet transits using a small telescope', *J. Brit. Astron. Assoc.*, **123**, 153–156 (2013)
13. Poddaný S., Brát L. & Pejcha O., 'Exoplanet Transit Database. Reduction and processing of the photometric data of exoplanet transits', *New Astron.*, **15**, 297–301 (2010)
14. Baluev R. V. *et al.*, 'Benchmarking the power of amateur observatories for TTV exoplanets detection', *Mon. Not. R. Astron. Soc.*, **450**, 3101–3113 (2015)
15. Baluev R. V. *et al.*, 'Homogeneously derived transit timings for 17 exoplanets and reassessed TTV trends for WASP-12 and WASP-4', *Mon. Not. R. Astron. Soc.*, **490**, 1294–1312 (2019)
16. Mallonn M. *et al.*, 'Ephemeris refinement of 21 hot Jupiter exoplanets with high timing uncertainties', *Astron. Astrophys.*, **622**, A81 (2019)
17. Zellem R. T. *et al.*, 'Engaging citizen scientists to keep transit times fresh and ensure the efficient use of transiting exoplanet characterization missions', *arXiv Prepr.*, **1903.07716**, 1–6 (2019)
18. Zellem R. T. *et al.*, 'Utilizing small telescopes operated by citizen scientists for transiting exoplanet follow-up', *Publ. Astron. Soc. Pacific*, **132**, 054401 (2020)
19. NASA Exoplanet Watch. Available at: **https://exoplanets.nasa.gov/exoplanet-watch/** (accessed 2020 April).
20. Kokori A. & Tsiaras A., 'ExoWorlds Spies: a project for public involvement in exoplanet research', *EPSC Abstr.*, **12**, EPSC2018-1268–1 (2018)
21. ARIEL ExoClock Project, by the ARIEL Ephemerides Working Group. Available at: **https://www.exoclock.space/** (accessed 2020 April).
22. Gould R. R., Sunbury S. & Krumhansl R., 'Using online telescopes to explore exoplanets from the physics classroom', *Am. J. Phys.*, **80**, 445–451 (2012)
23. Gould R., Sunbury S. & Dussault M., 'In praise of messy data: Lessons from the search for alien worlds', *Sci. Teach.*, 31–37 (2014)
24. NASA's Universe of Learning. Available at: **https://www.universe-of-learning.org/** (accessed 2020 April).
25. DIY Planet Search. Available at: **https://www.cfa.harvard.edu/smgphp/otherworlds/OE/** (accessed 2020 April).
26. Fowler M. J. F., 'Observing transiting 'hot Jupiter' exoplanets with the MicroObservatory', *J. Br. Astron. Assoc.*, **129**, 174–175 (2019)
27. Hartman J. D. *et al.*, 'HAT-P-32b and HAT-P-33b: Two highly inflated hot Jupiters transiting high-jitter stars', *Astrophys. J.*, **742**, 59 (2011)
28. Wang Y.-H. *et al.*, 'Transiting Exoplanet Monitoring Project (TEMP). V. Transit follow up for HAT-P-9b, HAT-P-32b, and HAT-P-36b', *Astron. J.*, **157**, 82 (2019)

**Table 4. Median values & 68% confidence intervals for HAT-P-32b system parameters**

| Parameter | Definition & units | This work | | Wang et al. (2019) | |
|---|---|---|---|---|---|
| **Stellar parameters:** | | | | | |
| $M_*$ | Mass ($M_\odot$) | 1.196 | +0.061 / −0.056 | 1.132 | +0.051 / −0.050 |
| $R_*$ | Radius ($R_\odot$) | 1.228 | +0.032 / −0.027 | 1.367 | +0.031 / −0.030 |
| $L_*$ | Luminosity ($L_\odot$) | 2.10 | +0.15 / −0.13 | 2.178 | +0.099 / −0.096 |
| $\rho_*$ | Density (cgs) | 0.917 | +0.034 / −0.052 | 0.625 | ±0.014 |
| $\log(g_*)$ | Surface gravity (cgs) | 4.338 | +0.013 / −0.017 | 4.22 | ±0.04 |
| $T_{eff}$ | Effective temperature (K) | 6274 | ±64 | 6001 | ±88 |
| [Fe/H] | Metallicity (dex) | −0.44 | +0.078 / −0.080 | −0.16 | ±0.08 |
| **Planetary parameters:** | | | | | |
| $P$ | Period (d) | 2.15000796 | ±0.00000069 | 2.15000820 | ±0.00000013 |
| $a$ | Semi-major axis (au) | 0.0346 | +0.00058 / −0.00054 | 0.03397 | +0.00051 / −0.00050 |
| $R_p$ | Radius ($R_J$) | 1.810 | +0.053 / −0.042 | 1.980 | ±0.045 |
| $T_{eq}$ | Equilibrium temp. (K) | 1802 | +24 / −22 | 1835.7 | +6.8 / −6.9 |
| $(F)$ | Incident flux ($10^9$ erg s$^{-1}$ cm$^{-2}$) | 2.39 | +0.13 / −0.11 | 2.505 | +0.042 / −0.052 |
| **Primary transit parameters** | | | | | |
| $T_C$ | Time of transit (BJD$_{TDB}$) | 2457692.75939 | +0.00031 / −0.00030 | 2455867.402743 | ±0.000049 |
| $R_P/R_*$ | Radius of planet in stellar radii | 0.1517 | ±0.0015 | 0.14886 | +0.00056 / −0.00054 |
| $a/R_*$ | Semi-major axis in stellar radii | 6.071 | +0.073 / −0.12 | 5.344 | +0.040 / −0.039 |
| $u_1$ | Linear limb-darkening coeff. | 0.287 | +0.042 / −0.041 | 0.316 | … |
| $u_2$ | Quadratic limb-darkening coeff. | 0.283 | +0.048 / −0.047 | 0.303 | … |
| $i$ | Inclination (degrees) | 88.8 | +0.83 / −1.1 | 88.98 | +0.68 / −0.85 |
| $b$ | Impact parameter | 0.128 | +0.11 / −0.09 | 0.083 | +0.070 / −0.055 |
| $\delta$ | Transit depth | 0.023 | ±0.0004 | 0.02216 | +0.00017 / −0.00016 |
| $T_{FWHM}$ | FWHM duration (d) | 0.11202 | +0.00097 / −0.00090 | 0.11284 | +0.00037 / −0.00038 |
| $\tau$ | Ingress/egress duration (d) | 0.01745 | +0.00085 / −0.00033 | 0.01712 | +0.00032 / −0.00015 |
| $T_{14}$ | Total duration (d) | 0.1297 | +0.0012 / −0.0011 | 0.13002 | +0.00052 / −0.00049 |
| $F_0$ | Baseline flux | 0.99873 | +0.00024 / −0.00025 | 0.99940 | ±0.00013 |
| **Secondary eclipse parameters:** | | | | | |
| $T_S$ | Time of eclipse (BJD$_{TDB}$) | 2457693.83440 | +0.00031 / −0.00030 | 2456236.268 | +0.10 / −0.098 |

29 Zhao M. *et al.*, 'Characterization of the atmosphere of the hot Jupiter HAT-P-32Ab and the M-dwarf companion HAT-P-32B', *Astrophys. J.*, **796**, 115 (2014)
30 Gibson N. P. *et al.*, 'The optical transmission spectrum of the hot Jupiter HAT-P-32b: clouds explain the absence of broad spectral features?', *Mon. Not. R. Astron. Soc.*, **436**, 2974–2988 (2013)
31 Nortmann L. *et al.*, 'The GTC exoplanet transit spectroscopy survey – IV. Confirmation of the flat transmission spectrum of HAT-P-32b'. *Astron. Astrophys.*, **594**, A65 (2016)
32 Mallonn M. & Strassmeier K. G., 'Transmission spectroscopy of HAT-P-32b with the LBT: confirmation of clouds/hazes in the planetary atmosphere', *Astron. Astrophys.*, **590**, A100 (2016)
33 Mallonn M. & Wakeford H. R., 'Near-ultraviolet transit photometry of HAT-P-32 b with the Large Binocular Telescope: Silicate aerosols in the planetary atmosphere', *Astron. Nachrichten*, **338**, 773–780 (2017)
34 Seeliger M. *et al.*, 'Transit timing analysis in the HAT-P-32 system'. *Mon. Not. R. Astron. Soc.*, **441**, 304–315 (2014)
35 Each of the five telescopes in the MicroObservatory network are light-heartedly named after noted astronomers: *Annie* Jump Cannon, *Ben*jamin Banneker, *Cecilia* Payne-Gaposchkin, *Donald* Menzel, and *Ed*ward Pickering.
36 *C-MUNIPACK*. Available at: http://c-munipack.sourceforge.net/ (accessed 2020 April).
37 Eastman J., Gaudi B. S. & Agol E., 'EXOFAST: A fast exoplanetary fitting suite in IDL', *Publ. Astron. Soc. Pacific*, **125**, 83–112 (2013)
38 Eastman J., Siverd R. & Gaudi B. S., 'Achieving better than 1 minute accuracy in the heliocentric and barycentric Julian dates', *Publ. Astron. Soc. Pacific*, **122**, 935–946 (2010)
39 *MEOW* pipeline demonstration. Available at: **https://www.youtube.com/watch?v=fqVZho3fOkU** (accessed 2020 April).
40 Mandel K. & Agol E., 'Analytic light curves for planetary transit searches', *Astrophys. J.*, **580**, L171–L175 (2002)
41 Ford E. B., 'Quantifying the uncertainty in the orbits of extrasolar planets', *Astron. J.*, **129**, 1706–1717 (2005)
42 Ford E. B., 'Improving the efficiency of Markov Chain Monte Carlo for analyzing the orbits of extrasolar planets', *Astrophys. J.*, **642**, 505–522 (2006)
43 Ter Braak C. J. F., 'A Markov Chain Monte Carlo version of the genetic algorithm Differential Evolution: Easy Bayesian computing for real parameter spaces', *Stat. Comput.*, **16**, 239–249 (2006)
44 Auvergne M. *et al.*, 'The CoRoT satellite in flight: description and performance', *Astron. Astrophys.*, **506**, 411 (2009)
45 Carter J. A. *et al.*, 'Analytic approximations for transit light curve observables and uncertainties', *Astrophys. J.*, **689**, 499–512 (2008)
46 Fowler M. J. F., 'Serendipitous observations of UCAC4 686-012519: a short period δ Scuti pulsating star in Andromeda', *Brit. Astron. Assoc. Variable Star Section Circ.*, no. 186, 33–38 (2020)

# New members

**2021 January 23**

ALEXANDER Vernon, Stirling
ANDERSON Robert, Renfrewshire
BARRATT Michael, Wiltshire
BEECH Rebecca, Surrey
BENNETT Gillian, Worcestershire
BEZ-CRYER Isabella, Bath & North East Somerset
BEZUGLAYA Lena, Hammersmith & Fulham
BODDINGTON David, Berkshire
CERAVOLO Peter, British Columbia, CANADA
COCCIA Francesco, Napoli, ITALY
DAVIES Alun, North Yorkshire
DAVIS Anne, Kent
DRENNEN Debbie, Devon
DUBEY Kamlesh, Melaka, MALAYSIA
FELTON Miles, Cumbria
GEORGE Jodie, Poole
GREEN Tony, Coventry
GRICE Keith, Shropshire
HAMMOND James, East Ayrshire
HOULTON Nigel, Derbyshire
HUMBERSTONE Joseph, Conwy
Institute de Astronomia, Birmingham, USA
JONES-WELLS Owen, Dorset
KERESZTY Zsolt
KNAGGS Michael, Leicestershire
LEE Mark, Pembrokeshire
LEWIS Roger, Cardiff
MADHAVAN Jaswant, Franklin, USA
MAROTTA Michael, Texas, USA
MORRIS Brian, JERSEY
MURRAY Paul, Berkshire
OMER Michael, Gloucestershire
PATEL Pinakin, Cambridgeshire
PICKFORD Chris, Oxfordshire
PISTRITTO Davide, Bari, ITALY
PLATTS Chris, East Sussex
PLUME Christopher
RECORD Geoff, Derby
ROBINSON James, Wiltshire
ROWE Ken, London
SHADRACH David, Alpes-Maritimes, FRANCE
SMITH-SUAREZ Edward, Southwark
STEPHEN Neil, Edinburgh
THOMSON Matthew, London
TRUNLEY Harold, Lancashire
VAMPLEW Anton, Kent
WADE Thomas, Rotherham
WAREING Mark, Cumbria
YORK Stephen, Leicestershire
YOUNG Gordon, Essex

**2021 March 31**

AKTAS Sezer, İstanbul, TURKEY
ANDERSON Colin, Berkshire
ANNE Susan, Pembrokeshire
BOWSKILL Grant, Bedfordshire
BUCHHEIM Robert
BUCKLEY Derek, Dublin, IRELAND
CLARK Russell, Neath Port Talbot
CONWAY Dr Andrew, Glasgow
COOKE Martin, Alava, SPAIN
DAVIES Karl, Kent
DAVIS Mark
DOWNIE Gordon, Bath & North East Somerset
ELDRIDGE Colin, Western Australia, AUSTRALIA
EVANS Peter, London
FENTAMAN Arthur, Kent
FERNANDES Rollando, Leicestershire
FORBES Hilary
GREENHILL-HOOPER Mike, Occitanie, FRANCE
HODSON Brian, Lancashire
HOFFMANN Brian, York
JACKSON Leslie, Derby
KIISKINEN Harri, Jyväskylä, FINLAND
KIRK Adam, Nottinghamshire
KLAGES Peter, Shropshire
MALAMATENIOS David, Hertfordshire
MARSHALL James, Lincolnshire
MARSHALL Riya, Lincolnshire
McDONALD Gary, London
McKEE Robert, Greater Manchester
MENZIES Kenneth, Massachusetts, USA
MILLS Peter, Norfolk
MORAN John, East Yorkshire
NEAL Mark, Berkshire
NELSON Peter, Liverpool
OWEN Kirsten, Hertfordshire
PARKES Dr Harry, Lewisham
PATTINSON David, Staffordshire
PEARSON Frank, Staffordshire
PEREZ SOLAR Christian, Angus
PYWELL Richard Francis, Cornwall
ROBERTSON Raymond, Brighton & Hove
ROMERO Luis, Barcelona, SPAIN
SILOW Fredrik, SWEDEN
SMITH Robert, Warwickshire
STONE Lynne, South Gloucestershire
STRICKLAND Stephen, West Yorkshire
TEMLETT Tom, Somerset
VALVASORI Adriano, Sala Bolognesa, ITALY
WARD Matthew, Dorset
WARDLEY Peter
WILBY Christopher, Bolton
WYKES Rory, Northamptonshire